\documentclass[conference]{IEEEtran}
\IEEEoverridecommandlockouts
\usepackage{cite}
\usepackage{amsmath,amssymb,amsfonts}
\usepackage{algorithmic}
\usepackage{graphicx}
\usepackage{textcomp}
 \usepackage{hyperref}
\usepackage{xcolor}
\def\BibTeX{{\rm B\kern-.05em{\sc i\kern-.025em b}\kern-.08em
    T\kern-.1667em\lower.7ex\hbox{E}\kern-.125emX}}
\begin{document}

\title{Ultrasound Visualization using VTK}

\author{\IEEEauthorblockN{Bhavya Sehgal}
\IEEEauthorblockA{\textit{Master of Multimedia} \\
\textit{University of Alberta}\\
Edmonton, Canada \\
Email: bsehgal1@ualberta.ca\\
}
\and
\IEEEauthorblockN{Gavin (Zichen) Gui}
\IEEEauthorblockA{\textit{Master of Multimedia} \\
\textit{University of Alberta}\\
Edmonton, Canada \\
Email: zgui@ualberta.ca\\
}
\and
\IEEEauthorblockN{Md Nahid Sadik}
\IEEEauthorblockA{\textit{Master of Multimedia} \\
\textit{University of Alberta}\\
Edmonton, Canada \\
Email: msadik@ualberta.ca\\
}
}

\maketitle

\begin{abstract}
This project developed a web application using VTK for ultrasound visualization. The images were enhanced using median and Gaussian filters, and two algorithms were utilized for data visualization: isosurface extraction and Delaunay triangulation. Results showed that both algorithms were effective at reducing Gaussian noise and high-frequency noise, such as speckle noise, which is common in ultrasound. The web application allows users to select MHA files, adjust the marching cubes threshold, and switch between the two algorithms in runtime. This project demonstrates the benefits of ultrasound visualization in medical applications and the utility of using VTK to achieve high-quality visualizations. The web application provides healthcare professionals with a user-friendly platform to interpret ultrasound data, leading to better diagnosis and treatment decisions.
\end{abstract}


\section{Introduction}
Medical imaging has made great strides with the use of tools such as ultrasound, MRI, and CT scans\cite{wee2011surface}. However, interpreting these images can still be a challenging task that requires specialized knowledge and software \cite{hafizah20103d}. To address this, a web application was developed using the Visualization Toolkit (VTK) for ultrasound visualization. The MHA file format, commonly used in medical imaging, was employed for the 3D volumetric data, and both median and Gaussian filters were utilized to enhance image quality. Two algorithms, isosurface extraction and Delaunay triangulation, were used for data visualization, and users can switch between the two algorithms in runtime to obtain different visualizations. The aim of the project is to provide healthcare professionals with a user-friendly and accessible platform to interpret ultrasound data and make better diagnosis and treatment decisions.\\

The interactive features of the web application allow users to explore the data and customize the visualization to their specific needs. By utilizing the power of VTK and the flexibility of web-based technologies, this project aims to improve the accessibility and quality of medical imaging data analysis. Ultimately, the goal is to make medical imaging technologies more accessible and user-friendly, providing healthcare professionals with better tools to interpret imaging data and improve patient outcomes.
\section{Related works}

\subsection{3D Reconstruction of 2D Medical Images From Dicoms Files}

This paper \cite{kumar3d} discusses the utilization of PACS in medical imaging and how it stores and retrieves DICOM information using RDBMS. The benefits of using PACS, such as research, treatment strategies, and disease progression monitoring, are also highlighted. Additionally, the paper introduces 3D techniques in medical imaging, including MPR, SR, and VR, and discusses how they can be used to visualize tumors and other objects. Image segmentation is also discussed as a method of breaking down digital images into subgroups for analysis. Overall, the paper provides an overview of how PACS and 3D imaging techniques can benefit medical professionals in treating patients.

\subsection{3D Digital Reconstruction of Brain Tumor From MRI Scans}

The paper\cite{bharathi20153d} proposes an algorithm for 3D reconstruction of brain tumors from MRI images involves segmentation, Delaunay triangulation, and triangular surface visualization. Pre-processing and segmentation involve noise elimination, histogram equalization, and elimination of the skull region. Mathematical morphology is used for image processing. The algorithm calculates the contour plot of the matrix Z using vectors x and y to control scaling on the x and y-axes, and the contours are superimposed in 3D to create a better 3D reconstruction of the brain tumor from MRI images. The algorithm shows promising results for tumor detection and could be used for diagnostic purposes.
\section{Methodology}
The methodology followed in the project involved several steps, including data processing, developing visualization techniques, building a web application, conducting a usability study, and deploying the application. These steps were critical in ensuring the effectiveness and usability of the ultrasound visualization techniques developed in the project.
\begin{enumerate}
  \item Obtaining the dataset: The ultrasound dataset was obtained from \url{https://www.vicomtech.org/demos/us_tracked_dataset/UsTrackedDataset.htm}. The dataset used in the project was the Baby Phantom Completed Sequence Ultrasound Dataset.
  
  \item Data processing: The dataset was processed using median and Gaussian filters to remove noise and improve image quality.
  
  \item Visualization techniques: Two different visualization techniques, isosurface extraction and Delaunay triangulation, were used to visualize the processed ultrasound data.

  \item Development of web application: The web application was developed that allows users to select the MHA file they want to visualize, adjust the marching cubes threshold, and switch between the two visualization techniques in real-time.

  \item Usability study: A usability study was conducted to evaluate the application's effectiveness and ease of use.

  \item Deployment: The application was deployed on a web server, making it accessible to users from anywhere with an internet connection.
\end{enumerate}

\section{Dataset}
\begin{figure}[ht]
    \includegraphics[width = 9cm]{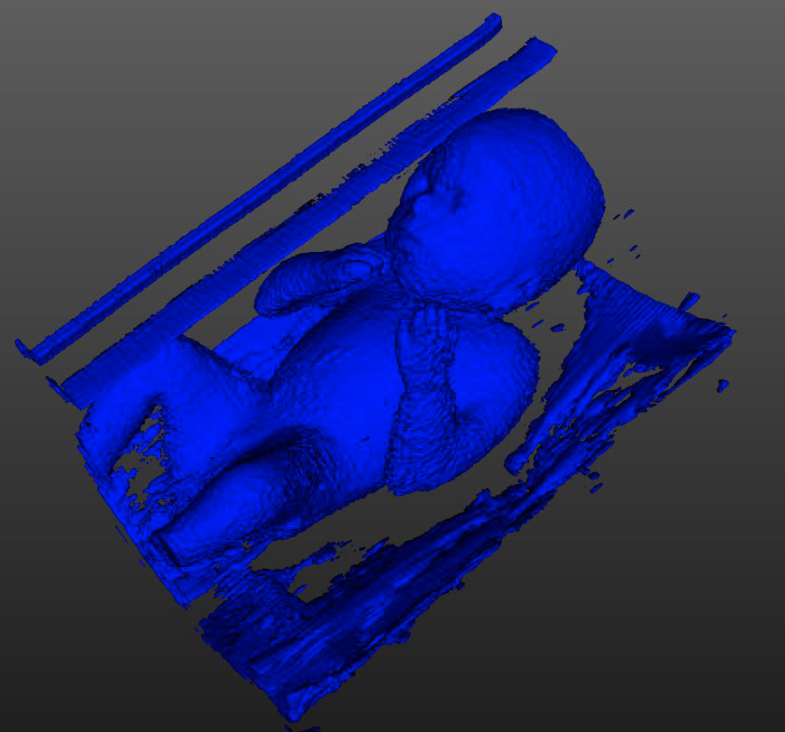}
    \caption{Baby Embryo dataset}
    \label{Dataset}
\end{figure}

For this project, we utilized the Baby Phantom Completed Sequence Ultrasound Dataset from the US Tracked Dataset provided by the Vicomtech research center\cite{cortes2016ultrasound} as shown in \ref{Dataset}.This dataset contains a series of ultrasound images captured using a clinical ultrasound machine, which were later processed and reconstructed using a tracking system to obtain accurate positional information.
The dataset was stored in the Meta Image Format (MHA), a file format commonly used in medical imaging, making it compatible with the visualization toolkit (VTK) library used in the project.The dataset consists of a series of 256 x 256 x 256 MHA files, which contain volumetric ultrasound data. The files are organized by different imaging planes, including sagittal, coronal, and axial views, providing different perspectives of the anatomy of the fetus. The ultrasound data in the dataset is affected by noise and artifacts, which can make interpretation challenging.
\section{Implementation Details}
\subsection{Data Pre-processing}
In this project, two different filters, median and Gaussian, were used to perform data preprocessing on ultrasound images. The median filter was applied to remove the noise from the ultrasound data, while the Gaussian filter was used to blur the images and reduce noise. The filters were applied using the Visualization Toolkit library, and the resulting preprocessed ultrasound data was saved in the Meta Image Format. The Gaussian filter performed well in reducing Gaussian noise and high-frequency noise, such as speckle noise, which is common in ultrasound images.\\

Data preprocessing using median and Gaussian filters was crucial in this project to improve the quality of the ultrasound images and enhance the accuracy of the subsequent visualization algorithms. By removing noise and smoothing out the image, the filters provided clearer and more accurate ultrasound data for visualization. The use of standard file formats such as MHA ensured that the preprocessed ultrasound data was easily accessible and compatible with other medical imaging software. Overall, data preprocessing was a vital step in the ultrasound visualization process and helped to ensure the accuracy and reliability of the final visualization results.\\
\subsection{Iso-Surface Extraction}
One of the main visualization techniques used in the project for ultrasound data was isosurface extraction, which involves generating a 3D surface from a 3D image dataset by selecting a threshold value and extracting all the voxels with a value above that threshold \cite{narkbuakaew20083d}. This technique was applied in conjunction with a Gaussian filter to enhance the image quality of the ultrasound data, resulting in a clear and detailed 3D view of the ultrasound data. The web application developed for the project allows users to adjust the marching cubes threshold used in the isosurface extraction technique to fine-tune the visualization to their specific needs, giving them more flexibility when visualizing ultrasound data.\\

\begin{figure}[ht]
    \includegraphics[width = 8cm]{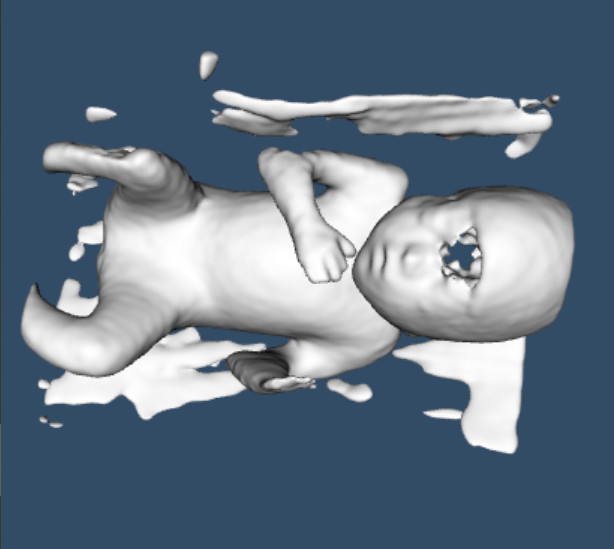}
    \caption{Isosurface Extraction}
    \label{Isosurface}
\end{figure}

The combination of isosurface extraction and the Gaussian filter proved to be an effective technique in enhancing the image quality of the ultrasound data. Isosurface extraction allowed for the generation of a 3D surface that provided a more detailed and accurate representation of the ultrasound data. The Gaussian filter helped to reduce noise and improve the overall quality of the images, resulting in a clearer visualization. The ability to adjust the marching cubes threshold in real-time through the web application provides users with the flexibility to optimize the visualization according to their preferences and specific needs.
\subsection{Delaunay Triangulation}
\begin{figure}[ht]
    \includegraphics[width = 8cm]{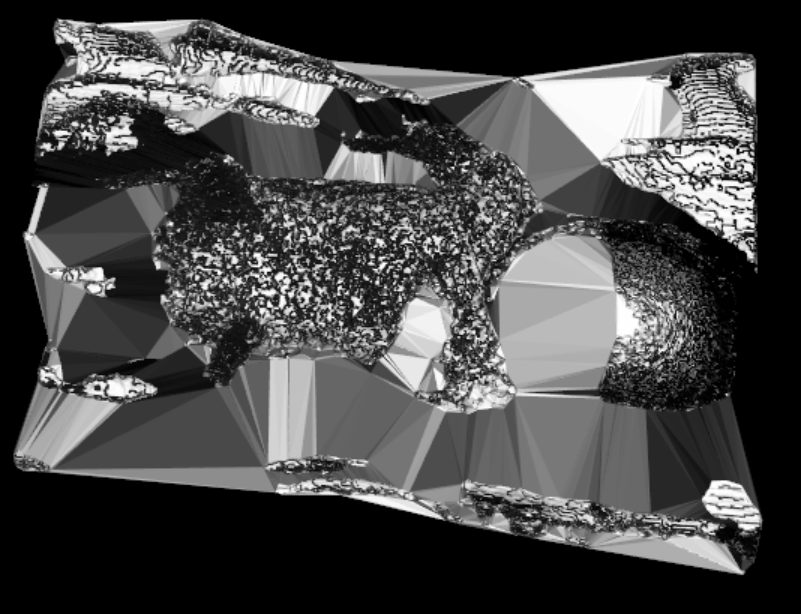}
    \caption{Delaunay Triangulation}
    \label{Delaunay}
\end{figure}

Delaunay triangulation 2D is another visualization technique used in the project to visualize ultrasound data \cite{lee1980two}. It is used to generate a set of non-overlapping triangles from a set of points. The resulting triangulation maximizes the minimum angle of all the triangles, providing a smoother and more regular visualization. This technique was used in conjunction with a Gaussian filter to enhance the image quality of ultrasound data as shown in \ref{Delaunay}.\\

The web application developed for the project includes the ability to switch between Delaunay triangulation 2D and isosurface extraction techniques in real-time, enabling users to compare and choose the best technique for their specific needs. The application also includes an adjustable marching cubes threshold, which allows users to fine-tune the visualization. Overall, the resulting visualization provides a clear and detailed 2D view of the ultrasound data.

\subsection{Development and Deployment of the Ultrasound Visualization Web Application}
TThe Ultrasound Visualization Web Application was developed to provide medical practitioners with an accessible platform for displaying ultrasound data. The online application is easily accessible from any device with an internet connection, without the need for additional software installation. Medical practitioners can upload their own Meta Image Formatted dataset and view it using the visualization techniques available in real-time via the user-friendly interface. This gives medical practitioners more flexibility in visualizing ultrasound data, enabling them to concentrate on certain areas of interest or spot potential problems that conventional ultrasound visualization approaches may have overlooked.\\

\begin{figure}[ht]
    \includegraphics[width = 9cm]{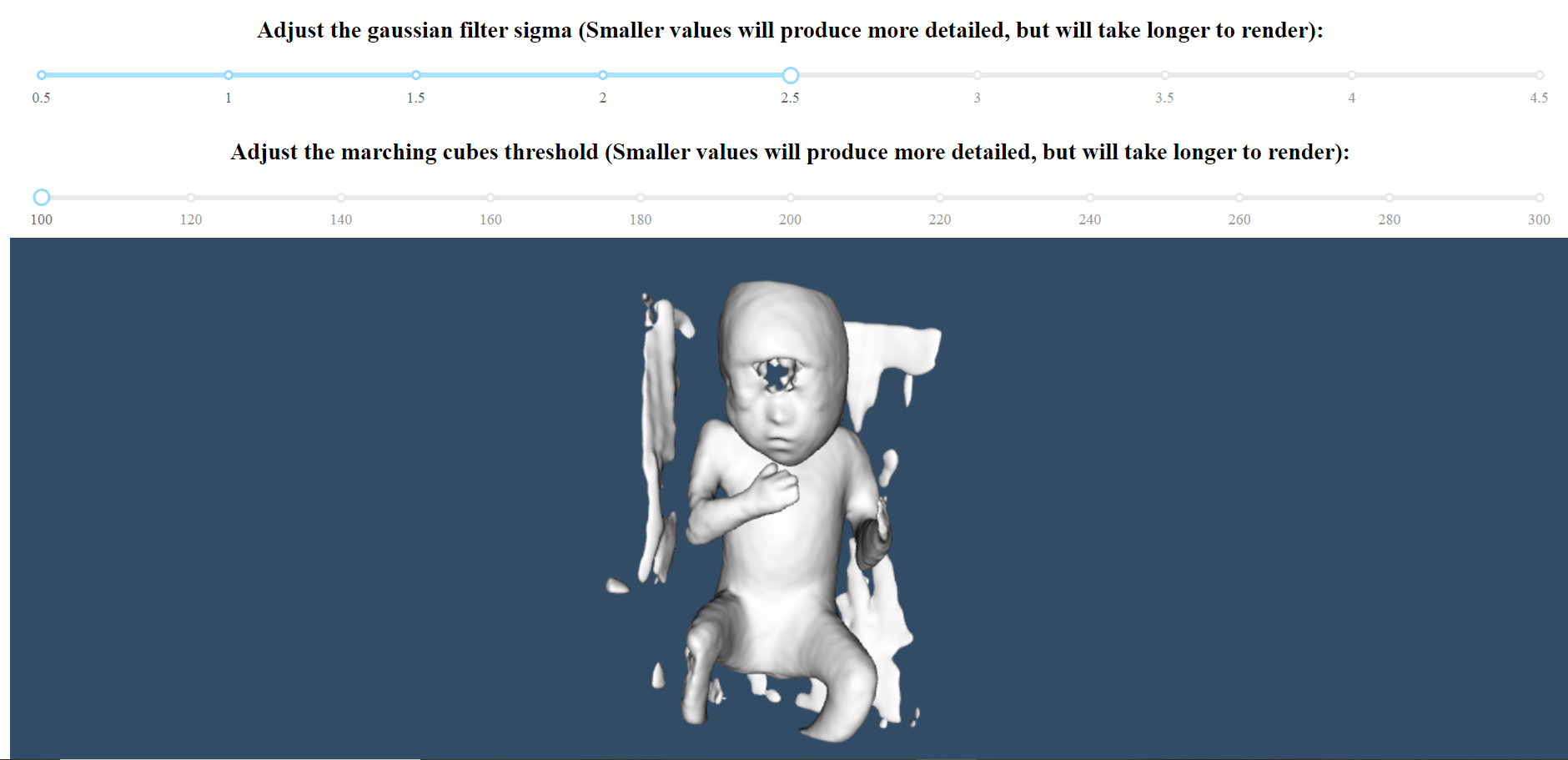}
    \caption{Ultrasound Visualization Web Application}
    \label{Website}
\end{figure}

The server-side application processes the MHA dataset uploaded by users to the website. The pre-processing of ultrasound data using a Gaussian filter increases the precision and reliability of the images. The data is then processed through the isosurface extraction and Delaunay triangulation algorithms, offering special anatomical insights. The web application presents a 3D representation of the ultrasound data after processing, which can be rotated and adjusted to highlight regions of interest. The user can switch between several visualization techniques, adjust the marching cubes threshold, and study the data in real-time to better understand the patient's anatomy.

\section{Results}
\subsection{Usability Testing}
The usability study of the ultrasound visualisation web application showed positive results with high average ratings across all evaluated areas. Participants found the application easy to learn and use, and consistent in behaviour. While there is room for improvement in terms of efficiency, the app's user-friendly design suggests its potential for increased popularity and utility in medical contexts. Further research could assess its effectiveness in specific medical situations for diagnosis and treatment planning.
\subsection{Medical Expert feedback}
The ultrasound visualization web application has received positive feedback from medical experts for its potential to improve the accuracy and reliability of ultrasound imaging. However, it's important to use the visualization alongside other diagnostic tools and medical expertise, and to ensure the accuracy of the ultrasound data obtained through appropriate protocols and equipment. Switching between isosurface extraction and Delaunay triangulation can provide unique insights into the patient's anatomy, highlighting surface features and offering a more detailed view of internal structures.
\section{Challenges}
There were several challenges faced during the development and implementation of the ultrasound visualization web application project. Some of the major challenges include:
\begin{enumerate}
    \item Data quality: Ensuring that the input ultrasound data was of high quality was one of the biggest challenges. The accuracy of the visualisation could be significantly impacted by any artefacts or inconsistencies in the data.
    \item Resource Requirements: To produce high-quality images, visualisation algorithms like Delaunay triangulation and marching cubes need a lot of computing power. For environments with limited resources, like mobile devices or low-power computers, this can be difficult.
    \item User Interface Design: Designing an intuitive user interface that enables medical professionals to easily manipulate and visualise ultrasound data was challenging.
    \item Performance optimization:  In order for the application to handle large datasets and offer real-time visualisation, it needed to be made more efficient.
\end{enumerate}
\section{Conclusion and Future Works}
In conclusion, this project has demonstrated the effectiveness of using VTK and web-based technologies to develop a user-friendly and accessible platform for ultrasound data visualization. The ability to adjust marching cubes threshold and switch between isosurface extraction and Delaunay triangulation algorithms provides healthcare professionals with greater flexibility and precision in interpreting ultrasound data. However, the accuracy and reliability of the ultrasound data is dependent on several variables, such as the patient's body habitus and the quality of the equipment used for data collection. It is essential to ensure that the appropriate techniques and equipment are used to collect high-quality input data to enhance the accuracy of the visualization.\\

In the future, the web application can be further enhanced by incorporating more sophisticated visualization algorithms and increasing its usability. It can also be extended to support other imaging techniques like CT and MRI scans. The integration of AI methods can also facilitate automated diagnosis and treatment planning. Overall, this project highlights the potential of web-based technologies to improve the accessibility and accuracy of medical imaging data analysis, which can lead to better diagnosis and treatment decisions for patients.
\bibliographystyle{IEEEtran}
\bibliography{bib.bib}
\end{document}